\documentclass[letterpaper]{article}

\usepackage{emulateapj}
\usepackage{onecolfloat}
\usepackage{apjfonts}
\usepackage{amsmath}
\usepackage{natbib}
\submitted{X-ray timing meeting proceedings}

\usepackage{graphicx}
\usepackage{float}       
\usepackage{afterpage}   

\newcommand{\rxte}{\emph{RXTE}}

\begin{document}

\twocolumn[

\title{Transitions of black hole transients to the low/hard state under the microscope}

\author{E. Kalemci\altaffilmark{1},
 J. A. Tomsick\altaffilmark{2},
 R. E. Rothschild\altaffilmark{2},
 K. Pottschmidt\altaffilmark{3,4},
 P. Kaaret\altaffilmark{5}
}

\altaffiltext{1}{Space Sciences Laboratory, 7 Gauss Way, University of
California, Berkeley, CA, 94720-7450, USA}

\altaffiltext{2}{Center for Astrophysics and Space Sciences, Code
0424, University of California at San Diego, La Jolla, CA,
92093-0424, USA}

\altaffiltext{3}{Max-Planck-Institut f\"ur Extraterrestrische Physik, Giessenbachstr. 1, 85748 Garching, Germany}

\altaffiltext{4}{\emph{INTEGRAL} Science Data Centre, Chemin d'\'Ecogia 16, 1290 Versoix, Switzerland}

\altaffiltext{5}{Harvard-Smithsonian Center for Astrophysics, 60 Garden Street,
 Cambridge, CA, 02138, USA}

\begin{abstract}
We characterized the evolution of spectral and temporal properties of several 
Galactic black hole transients observed between 1996-2001 using the data from 
well sampled PCA observations close to the transition to the low/hard state 
\cite{Kalemci_tez}. We showed that the changes in temporal properties are 
much sharper than the changes in the spectral properties, and it is much 
easier to identify a state transition with the temporal properties. The ratio 
of the power-law flux to the total flux in the 3-25 keV band increases close 
to the transition, and the power-law flux shows a sharp increase along with the
 changes in temporal properties during the transitions \citep{Kalemci03}. In 
this work we concentrate on the decay of two recent outbursts, from 
4U 1543$-$47, and H1743$-$322 and discuss the state transitions by tracking 
their daily, and sometimes hourly evolution, and interpret results based on the
 expectations from our earlier observations. 
\end{abstract}

\keywords{black hole physics -- X-rays:stars}
]

\section{Introduction}

\begin{figure}
   \includegraphics[height=.36\textheight]{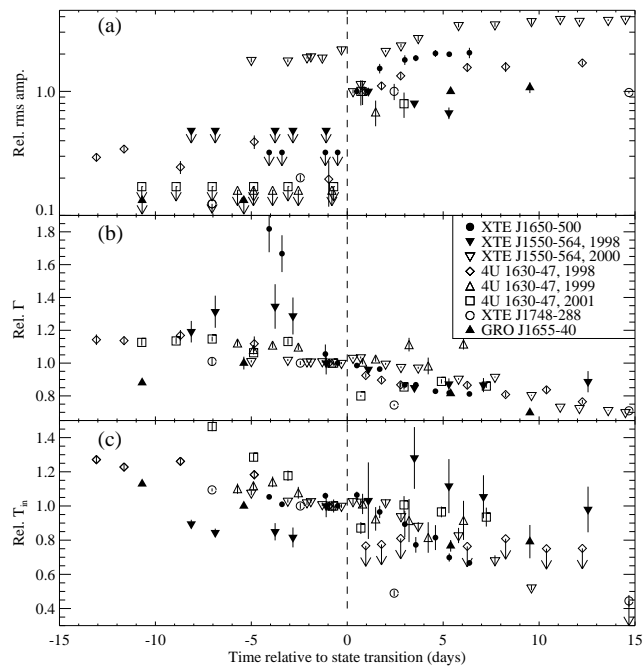}
   \caption{\label{fig:evol1}
Evolution of (a) rms amplitude of variability, (b) spectral index, and (c) 
inner disk temperature. The state transition is assumed to have happened in 
between the observations closest to the sharp change observed in panel (a), 
and represented by a dashed line. For (a), the values for each source are 
normalized with respect to the value just after the state transition. For 
both (b) and (c), the values for each source are normalized with respect to 
the value just before the state transition. 1 $\sigma$ errors.}
\end{figure}

\begin{figure}
   \includegraphics[height=.36\textheight]{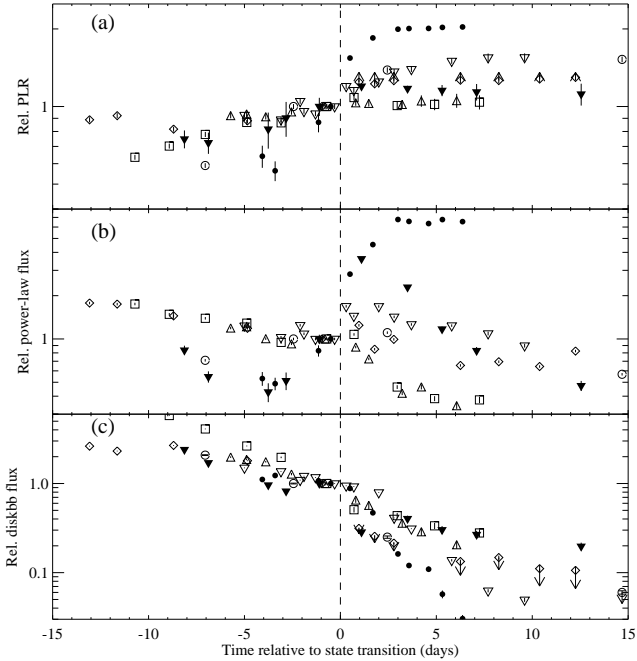}
   \caption{\label{fig:evol2}
Evolution of (a) the PLR, ratio of the power-law flux to the
total flux in 3-25 keV band, (b) the power-law flux, and (c) the diskbb 
flux. The dashed line represents the time of transition. The values for each 
source are normalized with respect to the value just before the state 
transition. 1 $\sigma$ errors.}
\end{figure}

The evolution of spectral and temporal properties for all Galactic Black Hole 
(GBH) transients that have been observed with \emph{RXTE} between 1996 and 
2001 that made a state transition during outburst decay was systematically 
studied in the dissertation of E. Kalemci \cite{Kalemci_tez}. For the spectral 
analysis, the data were fitted with a multi-component spectral model consisting
 of a power-law, a multi-color disk blackbody, a broad absorption edge 
 (\emph{smedge} model in XSPEC)with interstellar absorption. Power spectrum 
from each observation in 3 - 25 keV band was also created and fitted with 
Lorentzians. 5 sources in 8 outbursts were covered very well with \rxte\
 (close to daily monitoring), and the evolution of spectral and temporal 
properties from these outbursts can be summarized in Figs.~\ref{fig:evol1} 
and~\ref{fig:evol2}. During outburst decays, a very sharp change is observed 
in rms amplitude of variability, marking state transitions (shown with dashed 
lines in Figs.~\ref{fig:evol1}~and~\ref{fig:evol2}). The majority of the 
transitions are from a thermal dominant state (TD) to the hard state (HS),
however some sources goes through an intermediate state (IS) before 
reaching the HS (see \cite{McClintock03} for detailed definitions of these 
states). During the transitions, the evolutions of the spectral index, inner 
disk temperature and the diskbb flux are generally smooth, whereas a sharp 
change in power-law flux (Fig.~\ref{fig:evol2}.b) is usually observed along 
with a sharp change in temporal variability.


We observed two other GBH transients during outburst decay recently, 
4U 1543$-$47 in 2002 and H1743$-$322 in 2003. We applied a similar analysis
technique described above to see if these sources obey the general trends. In
addition to the daily 1-3 ks observations, we were also able to utilize a 20 ks
observation to characterize hourly evolution of a state transition of 
4U~1543$-$47.

\section{4U 1543$-$47, daily and hourly evolution}

\begin{figure}
  \includegraphics[height=.45\textheight]{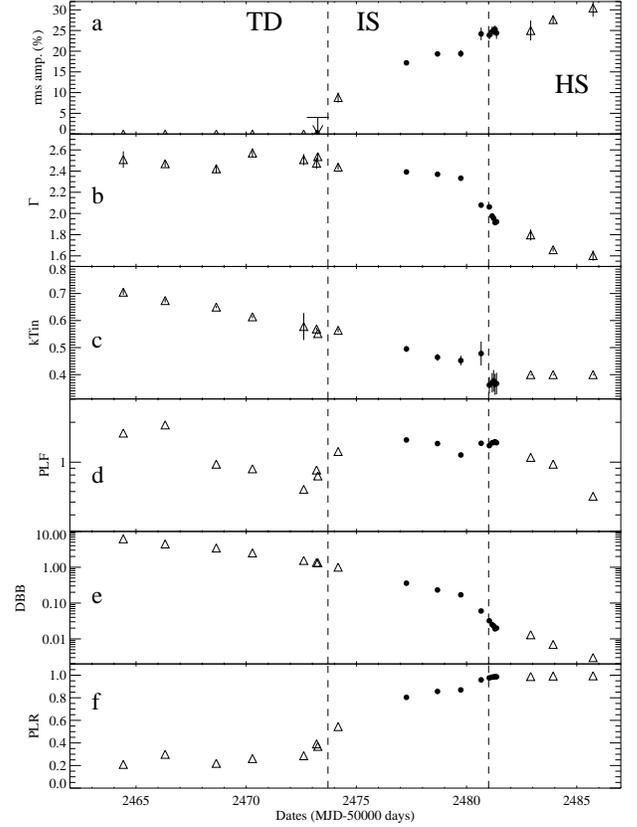}
  \caption{\label{fig:evol1543}
(a) Rms amplitude of variability, (b) spectral index, (c) 
inner disk temp., (d) power-law flux ($\rm 10^{-9}\; ergs\; cm^{-2}\, s^{-1}$),
(e) diskbb flux, (f) PLR. The dashed lines represent the approximate time of 
state transitions. The points represented by circles also show a QPO. These 
observations are also used in Fig.~\ref{fig:evol2}.
}
\end{figure}

The spectral and temporal evolution of 4U~1543$-$47 right around the transition
from TD to IS is shown in Fig.~\ref{fig:evol1543} (see also reports by 
M. Buxton and J. A. Tomsick on this source in this proceedings). Very weak 
temporal variability is observed right before the dashed line, but well 
defined variability appears after the dashed line. The appearance of weak 
variability is coincident with the increase in the power-law flux. The 
evolution close to the state transition is very similar to the other sources, 
as very smooth evolution is observed for the spectral index, disk temperature 
and disk flux, and a jump is observed in the power-law flux. Close to 
MJD~52480, another state transition is observed, this time to the HS. This 
transition happened right in the middle of our long 20 ks observation which 
enabled us to track the hourly spectral and temporal evolution of a black hole 
transient during a state transition.

\begin{figure}
  \includegraphics[height=.26\textheight]{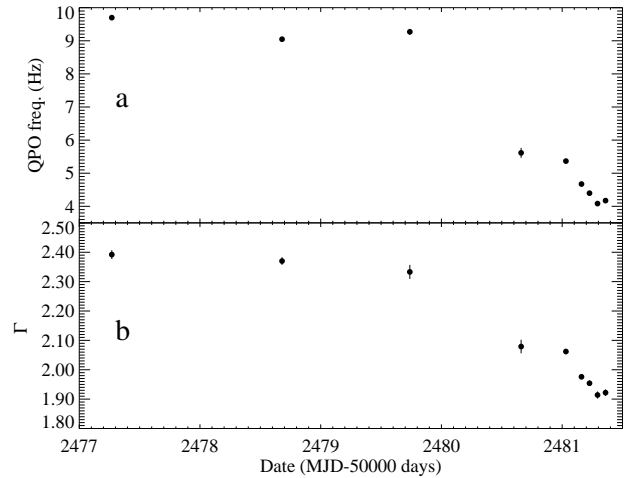}
  \caption{\label{fig:stevol1543}
(a) the QPO frequency (for the times when the QPO is observed), (b) the 
spectral index of 4U~1543$-$47 during the state transition.}
\end{figure}

During the second transition around MJD~52480, the inner disk temperature 
dropped sharply and the spectrum hardened. There was an increase in the 
power-law flux at the beginning which was accompanied by a jump in the rms 
amplitude of variability. The diskbb flux decreased an order of 
magnitude during this transition and finally the overall spectrum was dominated
by the power-law component, hence the source was in the HS. The short term 
evolution of the most interesting parameters of this transition, the QPO 
frequency and the spectral index are shown in Fig.~\ref{fig:stevol1543}. The 
last 6 pointings are only 2-3 hours apart. The QPO frequency follows the 
spectral index very tightly. The QPO frequency - spectral index correlation 
has been shown before (\cite{Kalemci_tez, Vignarca03}, see also Tomsick et al., this proceeding), but not in this kind of short time scales.

\clearpage

\section{Evolution of H1743-322}

\begin{figure}
  \includegraphics[width=0.35\textheight]{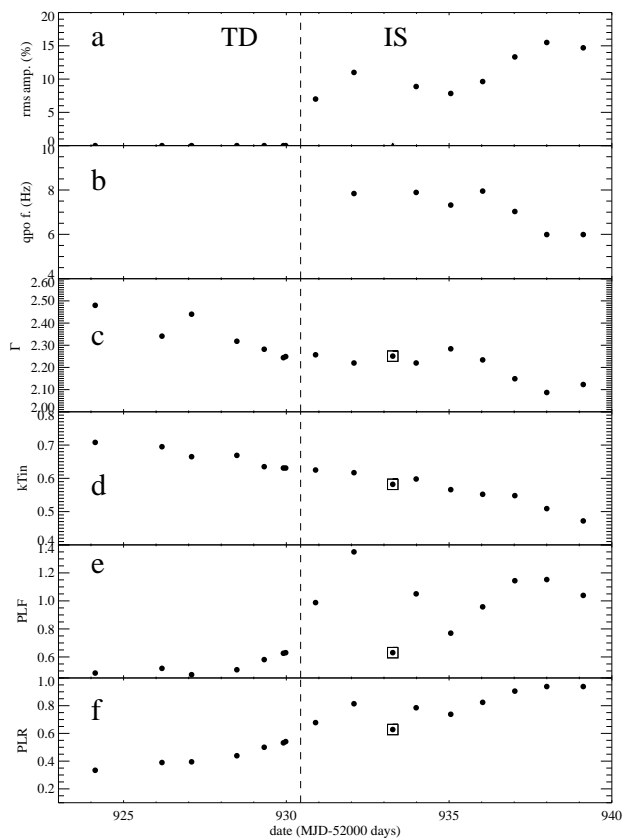}
  \caption{\label{fig:evol1743}
Evolution of spectral and temporal properties of H1743$-$322 during its decay 
of 2003 outburst (still in progress, \citealt{Tomsick03_atel}). (a) Rms 
amplitude of variability, (b) QPO frequency (if present) (c) spectral index, 
(d)inner disk temp., (e) power-law flux 
($\rm 10^{-9}\; ergs\; cm^{-2}\,s^{-1}$), (f) PLR. The dashed line represent 
the approximate time of state transitions. A second transition to the HS may 
have happened around MJD~52937. The observation shown with a square symbol 
showed no variability.
}
\end{figure}

The spectral and temporal evolution of H~1743$-$322 around the transition
from TD to IS is shown in Fig.~\ref{fig:evol1743}. Very similar to the other 
sources, and 4U~1543$-$47, the appearance of variability is coincident 
with a sudden increase of power-law flux. All other parameters during this 
transition shows a smooth evolution when the variability appeared. The 
correlation between the power-law index and the QPO frequency is, again, 
apparent. The observation at $\sim$MJD~52933 shows an interesting behavior 
(shown with a different symbol in Fig.~\ref{fig:evol1743}). The decrease in the
 power-law flux (and correspondingly the PLR) results in \emph{disappearance} 
of variability, although the transition has happened. This indicates a tight 
threshold for the PLR for appearance of variability.

\section{Discussion}
The uniform analysis of the evolution of the spectral and temporal properties
of several Galactic black hole transients during outburst decay expanded our
understanding of the state transitions and creation of variability 
\citep{Kalemci03}. It is now clear that the appearance of variability is 
related to an increase in the  power-law flux, and the rms amplitude of 
variability is related to the PLR. Both of the new outbursts that we analyzed 
in  light of our previous results supported these arguments. A very strong 
correlation between the QPO frequency and the power-law index was found after 
our analysis of BH transients \citep{Kalemci_tez}. Here, we also show how 
tightly the QPO frequency follows the spectral index, even for a few hours 
timescales for 4U~1543$-$47. This correlation may be interpreted as both 
parameters being a function of the position of the inner edge of the disk. If 
the accretion disk is close to the BH, perhaps a larger overlap between the 
corona and the disk occurs, resulting stronger cooling and softer spectrum 
\citep{Zdziarski02_2}. If the QPO frequency is related to a function of 
fundamental frequencies related to the inner edge (see for example 
\cite{Psaltis00}), these two parameters may be related. It is interesting
that such relation (if it is the correct interpretation) holds for
timescales as short as 2 hours.

\begin{figure}
  \includegraphics[width=0.35\textheight]{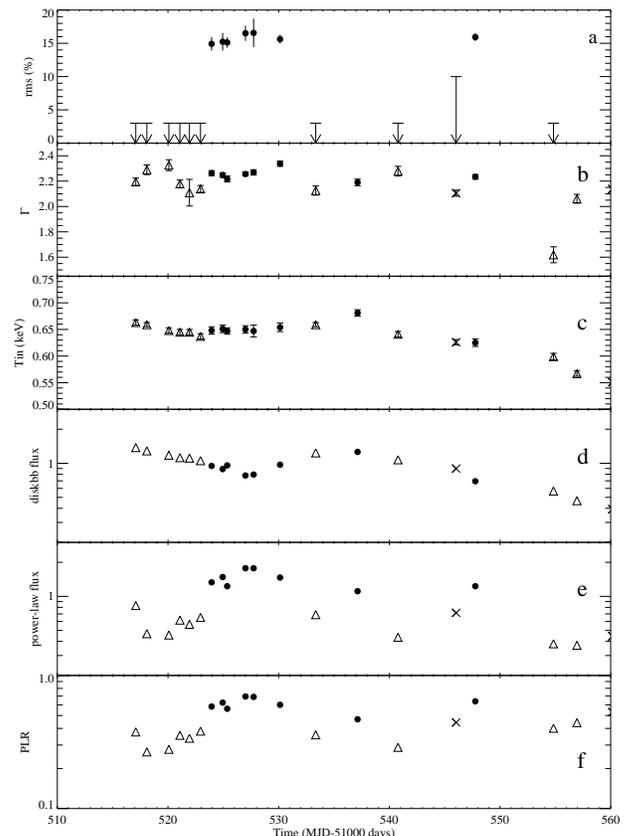}
  \caption{\label{fig:evol1859}
Spectral and temporal evolution of XTE~J1859$+$226 during its outburst decay 
in 2000. (a) Rms amplitude of variability, (b) spectral index, (c) 
inner disk temp., (d) diskbb flux ($\rm 10^{-9}\; ergs \; cm^{-2}\,s^{-1}$),
(e) power-law flux (same units), (f) PLR. Filled circles are used for 
observations with well defined variability with rms amplitude greater than 
15\%. The observation represented by the cross has rms amplitude less than 
10\%, and the remaining represented by triangles have rms amplitudes less 
than 3\%. 
}
\end{figure}

The disappearance of variability in H~1743$-$322 with a decrease in the PLR on 
MJD~52933 is very interesting, but not surprising for us. The same behavior, 
indicating a threshold PLR for variability, was also observed during the decay
 of XTE~J1859$+$226 in 2000 \citep{Kalemci03}. Fig.~\ref{fig:evol1859}
 illustrates the relation between spectral parameters and appearance of 
variability in XTE~J1859$+$226. The only parameters that change considerably 
when the variability appeared are the power-law flux and the PLR. For this
 source, the variability was observed when the PLR was greater than 0.45, and 
it disappeared when it was below this threshold value \citep{Kalemci_tez}. This
 threshold in the PLR may indicate a threshold size for the Comptonizing corona
 for state transitions to occur. It was argued that the mass accretion rate 
cannot be the ``only'' parameter that determines the spectral states of black 
hole systems, and a second independent parameter is required. 
Our observations support the idea that this second independent parameter is the
 size of the Comptonizing region as suggested earlier by Homan et 
al\citep{Homan01}.

\acknowledgments
E.K. acknowledges useful discussions with David Smith. Some part of this 
work was done at the Center for Astrophysics and Space Sciences (CASS) at UCSD 
as part of E.K.'s dissertation. E.K. acknowledges partial support 
of T\"UB\.ITAK. J.A.T. acknowledges partial support from NASA grant
NAG5-13055. K.P. was supported by grant Sta 173/25-1 and Sta 173/25-3 of the
Deutsche Forschungsgemeinschaft. P.K. acknowledges partial support from NASA 
grant NAG5-7405. This work was also supported by NASA contract NAS5-30720.


\begin{thebibliography}{}

\bibitem[\protect\astroncite{{Kalemci}}{2002}]{Kalemci_tez}
{Kalemci}, E.,  2002, \emph{Ph.D. Thesis}, University of California, San Diego


\bibitem[\protect\astroncite{{Kalemci} et~al.}{2003}]{Kalemci03}
{Kalemci}, E., {Tomsick}, J.~A., {Rothschild}, R.~E., {Pottschmidt}, K., and
  {Kaaret}, P., 2003, ApJ, astro-ph/0309799, accepted 

\bibitem[\protect\astroncite{{McClintock} \& {Remillard}}{2003}]{McClintock03}
{McClintock}, J.~E., \& {Remillard}, R.~A.,  2003,
\newblock in {X-ray Binaries, astro-ph/0306213}

\bibitem[\protect\astroncite{{Vignarca} et~al.}{2003}]{Vignarca03}
{Vignarca}, F., {Migliari}, S., {Belloni}, T., {Psaltis}, D., and {van der
  Klis}, M., 2003, A\&A, 397, 729--738 

\bibitem[\protect\astroncite{{Tomsick} and {Kalemci}}{2003}]{Tomsick03_atel}
{Tomsick}, J.~A., and {Kalemci}, E., 2003, \emph{ATEL}, 198

\bibitem[\protect\astroncite{{Zdziarski} et~al.}{2002}]{Zdziarski02_2}
{Zdziarski}, A.~A., {Poutanen}, J., {Paciesas}, W.~S., \& {Wen}, L.,  2002,
  ApJ, 578, 357

\bibitem[\protect\astroncite{{Psaltis} and {Norman}}{2000}]{Psaltis00}
{Psaltis}, D., and {Norman}, C., 2000, ApJ, submitted,
  astro-ph/0001391 

\bibitem[\protect\astroncite{{Homan} et~al.}{2001}]{Homan01}
{Homan}, J., {Wijnands}, R., {van der Klis}, M., {Belloni}, T., {van Paradjs},
  J., {Klein-Wolt}, M., {Fender}, R., \& {M{\'e}ndez}, M.,  2001, ApJS, 13,
  377


\end{thebibliography}

\end{document}